\documentclass[conference]{IEEEtran}
\IEEEoverridecommandlockouts
\usepackage{cite}
\usepackage{amsmath,amssymb,amsfonts}
\usepackage{algorithmic}
\usepackage{graphicx}
\usepackage{textcomp}
\usepackage{xcolor}
\def\BibTeX{{\rm B\kern-.05em{\sc i\kern-.025em b}\kern-.08em
    T\kern-.1667em\lower.7ex\hbox{E}\kern-.125emX}}
    
\usepackage{float}
\usepackage{subfig}

\usepackage{pgfplots}
\usepackage{tkz-euclide}
\usepackage{dblfloatfix}  

\usepackage{multirow}

\usetikzlibrary{shapes,arrows}
\usetikzlibrary{positioning}
\usetikzlibrary{chains}
\usetikzlibrary{intersections,calc}
\usetikzlibrary{automata}
\usetikzlibrary{backgrounds}
\usetikzlibrary{plotmarks}
\usetikzlibrary{decorations.markings}
\usetikzlibrary{patterns}
\pgfplotsset{compat=1.9}
    
\newlength\widthPlot
\newlength\heightPlot
\def\opacityPlot{0.1}
\def\scaleMarks{0.2}
\def\opacityMarks{.5}
\def\xminPlot{500}
\def\xmaxPlot{6000}
\def\tikzlegensPosY{-0.5in} 

\def\scaleGMCplotW{0.5}
\def\scaleGMCplotH{0.45}

\def\scaleVisuCompFig{0.3025}

\newcommand{\mat}{\boldsymbol}
\renewcommand{\vec}{\boldsymbol}   
    
\begin{document}

\title{Sparse Video Representation Using Steered Mixture-of-Experts With Global Motion Compensation}


\author{\IEEEauthorblockN{ Rolf Jongebloed, Erik Bochinski, and Thomas Sikora}
\IEEEauthorblockA{Communication Systems Group\\
Technische Universit\"at Berlin\\
Berlin, Germany\\}
}

\maketitle

\begin{abstract}
Steered-Mixtures-of Experts (SMoE) present a unified framework for sparse representation and compression of image data with arbitrary dimensionality.
Recent work has shown great improvements in the performance of such models for image and light-field representation.
However, for the case of videos the straight-forward application yields limited success as the SMoE framework leads to a piece-wise linear representation of the underlying imagery which is disrupted by nonlinear motion.
We incorporate a global motion model into the SMoE framework which allows for higher temporal steering of the kernels.
This drastically increases its capabilities to exploit correlations between adjacent frames by only adding $2$ to $8$ motion parameters per frame to the model but decreasing the required amount of kernels on average by $54.25\%$, respectively, while maintaining the same reconstruction quality yielding higher compression gains.
\end{abstract}


\section{Introduction}
\label{sec:intro}
In the last couple of years the potential of the \emph{Steered Mixture-of-Experts} (SMoE) framework for sparsely representing and coding images \cite{DBLP:conf/icip/VerhackSLWL16,8456250,dcc_smoe}, videos \cite{Videosmoe} and even higher dimensional imagery, i.e., light field images \cite{8019442} and light field videos \cite{smoe_lightfieldvideo,verhack_Journal} has been investigated.
As SMoE models give a description of the underlying data in the spatial domain instead using the frequency transform domain this compression approach drastically departs from conventional block-based coding techniques. 
As an extension of the well-known Mixture-of-Experts (MoE) approach it follows the divide-and-conquer principle \cite{6215056}.
Weighted Gaussian modes, also called kernels, collaborate in softmax gating functions, each indicating in which area its corresponding expert is trustworthy.
These Gaussians are characterized by their center positions and bandwidths within the pixel domain. 
This arrives at a soft partitioning of the input space with arbitrarily-shaped regions in which each expert acts as a regressor.
Such regions can be extended over the entire input space such that experts are responsible for thousands of pixels \cite{dcc_smoe}, allowing for efficient representation of the image data.\\
As the SMoE framework is easily scalable towards higher dimensional image modalities, in \cite{Videosmoe,8019442,smoe_lightfieldvideo,verhack_Journal}  this approach has been extended to video, light field image and light field video processing and coding, respectively.
They have shown that promising compression results can be achieved compared to state-of-the-art coding standards.
For the case of light field video which contains a 5D coordinate space (2 spatial, 2 angular and one temporal dimensions) SMoE already outperforms the well-elaborated multi-view high efficiency video coding standard (MV-HEVC) \cite{mvhevc}.
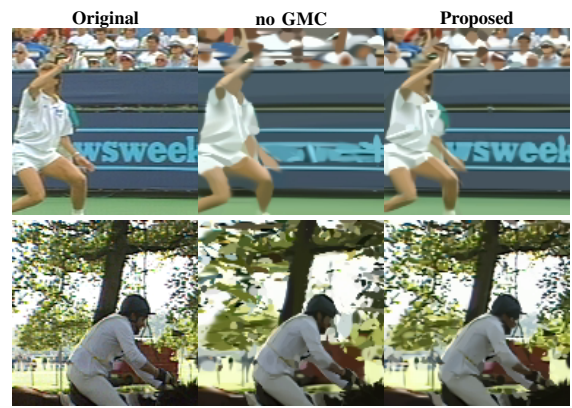
\begin{figure}[t]
\centering

\begin{tikzpicture}

\node[align=center, font=\bfseries, text=black,scale=.7]
at (0.45\textwidth * \scaleVisuCompFig * .5, 0.45\textwidth * \scaleVisuCompFig + 3.5pt) {Original};

\node[align=center, font=\bfseries, text=black,scale=.7]
at (0.45\textwidth * \scaleVisuCompFig * 1.5, 0.45\textwidth * \scaleVisuCompFig + 3.5pt) {no GMC};

\node[align=center, font=\bfseries, text=black,scale=.7]
at (0.45\textwidth * \scaleVisuCompFig * 2.5, 0.45\textwidth * \scaleVisuCompFig + 3.5pt) {Proposed};


\begin{axis}[%
width=0.45\textwidth * \scaleVisuCompFig,
height=0.45\textwidth * \scaleVisuCompFig,
at={(0.0in,0.0in)},
scale only axis,
xmin=0,
xmax=128,
y dir=reverse,
ymin=0,
ymax=128,
axis background/.style={fill=white},
axis line style={draw=none},
xticklabel=\empty,
yticklabel=\empty,
xtick=\empty,
ytick=\empty,
]
\addplot [forget plot] graphics 
[xmin=0,xmax=128,ymin=0,ymax=128]{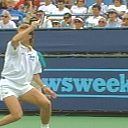};

\end{axis}


\begin{axis}[%
width=0.45\textwidth * \scaleVisuCompFig,
height=0.45\textwidth * \scaleVisuCompFig,
at={(0.45\textwidth * \scaleVisuCompFig,0.0in)},
scale only axis,
xmin=0,
xmax=128,
y dir=reverse,
ymin=0,
ymax=128,
axis background/.style={fill=white},
axis line style={draw=none},
xticklabel=\empty,
yticklabel=\empty,
xtick=\empty,
ytick=\empty,
]
\addplot [forget plot] graphics 
[xmin=0,xmax=128,ymin=0,ymax=128]{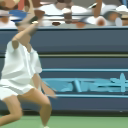};

\end{axis}


\begin{axis}[%
width=0.45\textwidth * \scaleVisuCompFig,
height=0.45\textwidth * \scaleVisuCompFig,
at={(0.45\textwidth * \scaleVisuCompFig * 2,0.0in)},
scale only axis,
xmin=0,
xmax=128,
y dir=reverse,
ymin=0,
ymax=128,
axis background/.style={fill=white},
axis line style={draw=none},
xticklabel=\empty,
yticklabel=\empty,
xtick=\empty,
ytick=\empty,
]
\addplot [forget plot] graphics 
[xmin=0,xmax=128,ymin=0,ymax=128]{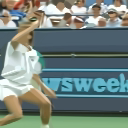};

\end{axis}


\begin{axis}[%
width=0.45\textwidth * \scaleVisuCompFig,
height=0.45\textwidth * \scaleVisuCompFig,
at={(0.0in,-0.45\textwidth * (\scaleVisuCompFig + 0.01))},
scale only axis,
xmin=0,
xmax=128,
y dir=reverse,
ymin=0,
ymax=128,
axis background/.style={fill=white},
axis line style={draw=none},
xticklabel=\empty,
yticklabel=\empty,
xtick=\empty,
ytick=\empty,
]
\addplot [forget plot] graphics 
[xmin=0,xmax=128,ymin=0,ymax=128]{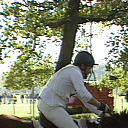};

\end{axis}


\begin{axis}[%
width=0.45\textwidth * \scaleVisuCompFig,
height=0.45\textwidth * \scaleVisuCompFig,
at={(0.45\textwidth * \scaleVisuCompFig,-0.45\textwidth * (\scaleVisuCompFig + 0.01))},
scale only axis,
xmin=0,
xmax=128,
y dir=reverse,
ymin=0,
ymax=128,
axis background/.style={fill=white},
axis line style={draw=none},
xticklabel=\empty,
yticklabel=\empty,
xtick=\empty,
ytick=\empty,
]
\addplot [forget plot] graphics 
[xmin=0,xmax=128,ymin=0,ymax=128]{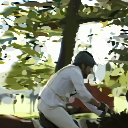};

\end{axis}


\begin{axis}[%
width=0.45\textwidth * \scaleVisuCompFig,
height=0.45\textwidth * \scaleVisuCompFig,
at={(0.45\textwidth * \scaleVisuCompFig * 2,-0.45\textwidth * (\scaleVisuCompFig + 0.01))},
scale only axis,
xmin=0,
xmax=128,
y dir=reverse,
ymin=0,
ymax=128,
axis background/.style={fill=white},
axis line style={draw=none},
xticklabel=\empty,
yticklabel=\empty,
xtick=\empty,
ytick=\empty,
]
\addplot [forget plot] graphics 
[xmin=0,xmax=128,ymin=0,ymax=128]{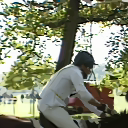};

\end{axis}

\end{tikzpicture}
\caption{Visual Comparison of frames at same number of kernels for \emph{Stefan} ($K \approx 2000$) and \emph{RaceHorse} ($K \approx  5000$) between SMoE without any GMC and proposed method}
\label{fig:visual_comparison}
\end{figure}
While MV-HEVC lacks in finding the optimal prediction structure, in SMoE each dimension is considered intrinsically at the same time by 5D Gaussian modes steering in its highest correlation to exploit as many redundancies as possible within the underlying input space.
The higher the dimensionality of the data the more redundancies emerge within, which is utilized naturally to the advantage of SMoE.\\
Nevertheless, these contributions for higher dimensional inputs still suffer from the usage of the Expectation-Maximization (EM) algorithm to train Gaussian Mixture Models jointly representing the experts and associated gates towards a maximized joint likelihood function which is not necessarily optimal for regression tasks. 
In \cite{8456250} and \cite{dcc_smoe} the EM algorithm has been replaced by the Gradient Descent (GD) to minimize directly the Mean Squared Error (MSE) or to maximize the structural similarity (SSIM) index of the reconstructed imagery, respectively. 
In \cite{dcc_smoe} the performance of image coding has been massively increased by also including the sparsification approach from \cite{tfsmoe} and introducing the model estimation with already quantized parameters compared to the initial work from \cite{DBLP:conf/icip/VerhackSLWL16}. 
Presumably, there are still improvements expected for the aforementioned higher dimensional imageries by applying these methods. 
However, this paper shows that the reconstruction quality suffers from non-linear camera motions within video sequences. 
Although the softmax gating introduces non-linearities each kernel has only linear steering capabilities \cite{verhack_Journal}
, and thus, a high number of kernels is necessary to compensate such motions between adjacent frames even though almost the same content is shown.
This contradicts the overall goal of obtaining sparse models with few kernels for efficient coding.
It is shown that Global Motion Compensation (GMC) can be applied to greatly reduce the number of required kernels at the same reconstruction quality for video data. 
As SMoE considers the pixels of the underlying imagery as data samples within a continuous space, each pixel can be displaced by perspective transformations frame by frame without introducing loss due to image warping. 
As a result, same content is spatially aligned and allows for higher temporal steering of the kernels, leading to a higher degree of sparsity of the final models. 
GMC has been used in past video codecs providing significant bitrate savings \cite{mpeg4-part2_asp} and also newer standards such as AVC and HEVC can still benefit through it  \cite{ryu2016improvement, tokPMM}.\\ 
This paper focuses on the modeling and sparse representation of video data.
The main contribution of this work will be the extension of the SMoE model by integration the temporal dimension and GMC for video sequences using the framework presented in \cite{tfsmoe}.
Evaluations show that the proposed method achieves the same reconstruction quality while decreasing the required amount of kernels by $54.25 \%$ compared to models without any motion compensation. Please note that our goal in this work is the extension of our framework to obtain sparse models of video sequences with high reconstruction qualities for future works (see. Fig.\ \ref{fig:visual_comparison}) rather than present a full video codec. Furthermore, the proposed method can be easily extended to light field video representations.

\section{Steered Mixture-of-Experts}

 We will review the Steered Mixture-of-Experts (SMoE) framework as follows \cite{dcc_smoe}:\\
%
The prediction function of the amplitudes $\vec{y}_p(\vec{x}) \in \mathbb{R}^3$ of a pixel (one luminance and two chrominance outputs) given its position $\vec{x} \in \mathbb{R}^d$ (spatial input; $d=3$ for video case) is formulated as a weighted sum of $K$ experts:
%
\begin{align} \label{eq:regression}
    \vec{y}_p(\vec{x}) = \sum\limits_{k=1}^K \vec{m}_{k}(\vec{x}) w_k(\vec{x}) \text{.}
\end{align}
In general, the experts  can be arbitrary regression functions, i.e.\ (hyper-)planes, polynomials, etc.,\ in the pixel domain, also called input space for each output channel. We choose them as hyperplanes:
%
  \begin{align} \label{eq:expert}
    \vec{m}_k\left( \vec{x} \right) &= \mat{M}_{k} \vec{x} + \vec{m}_{0,k}
\end{align}
with
\begin{align}
    \mat{M}_{k} = \left[\vec{m}_{\text{Y},k} \; \vec{m}_{\text{U},k} \; \vec{m}_{\text{V},k} \right]^T
\end{align}
where $\vec{m}_{\text{Y},k}$, $\vec{m}_{\text{U},k}$, and $\vec{m}_{\text{V},k}$ contain the slopes in the Y,U and V domains, respectively,
and $\vec{m}_{0,k}$ the offsets for each channel.
%
The weighting function, also called gating, in Eq.\ \ref{eq:regression} is a weighted soft max function
\begin{align}
w_k\left(\vec{x}\right)
&=
\frac{
	\pi_k\cdot \mathcal{K}\left(
	\vec{x};\vec{\mu}_{k},\mat{A}_{k}
	\right)
}{
\sum\limits_{j=1}^{K}
\pi_j\cdot \mathcal{K}\left(
\vec{x};\vec{\mu}_{j},\mat{A}_{j}
\right)
} \label{eq:softmax}
\end{align}
with mixing coefficients $\pi_k$. We employ Gaussian kernels 
\begin{align}
	\mathcal{K}\left(
	\vec{x};\vec{\mu},\mat{A} 
	\right)
	= \exp \left[
	-\frac{1}{2} (\vec{x} - \vec{\mu})^T \mat{A} \mat{A}^T(\vec{x} - \vec{\mu})
	\right]
\end{align}
defined by their center positions $\vec{\mu} \in \mathbb{R}^d$ and steering parameters \looseness=-1
   \begin{align}
   \mat{A}_k&:=
     \begin{pmatrix}
       a_{11,k} & \multicolumn{2}{c}{\kern0.5em\smash{\raisebox{-2.5ex}{\huge0}}} \\
       \vdots & \ddots &  \\
        a_{d1,k} & \hdots & a_{dd,k}
     \end{pmatrix}
   \end{align}
as the inverse cholesky decomposition of a covariance matrix where only the lower triangular part is nonzero and $d$ is the number of dimensions of the input space. 
Thus, it is guaranteed that the resulting matrix $\mat{\Sigma}^{-1} = \mat{A}\mat{A}^T$ is always positive semi-definite.\\
%
%
\begin{figure}[tb]
\centering
\begin{tikzpicture}

\begin{axis}[%
width=0.45\textwidth,
height=0.45\textwidth * 0.466,
at={(0.0in,0.0in)},
scale only axis,
xmin=0,
xmax=2253,
y dir=reverse,
ymin=0,
ymax=1050,
axis background/.style={fill=white},
axis line style={draw=none},
xticklabel=\empty,
yticklabel=\empty,
xtick=\empty,
ytick=\empty,
]
\addplot [forget plot] graphics 
[xmin=0,xmax=2253,ymin=0,ymax=1050]{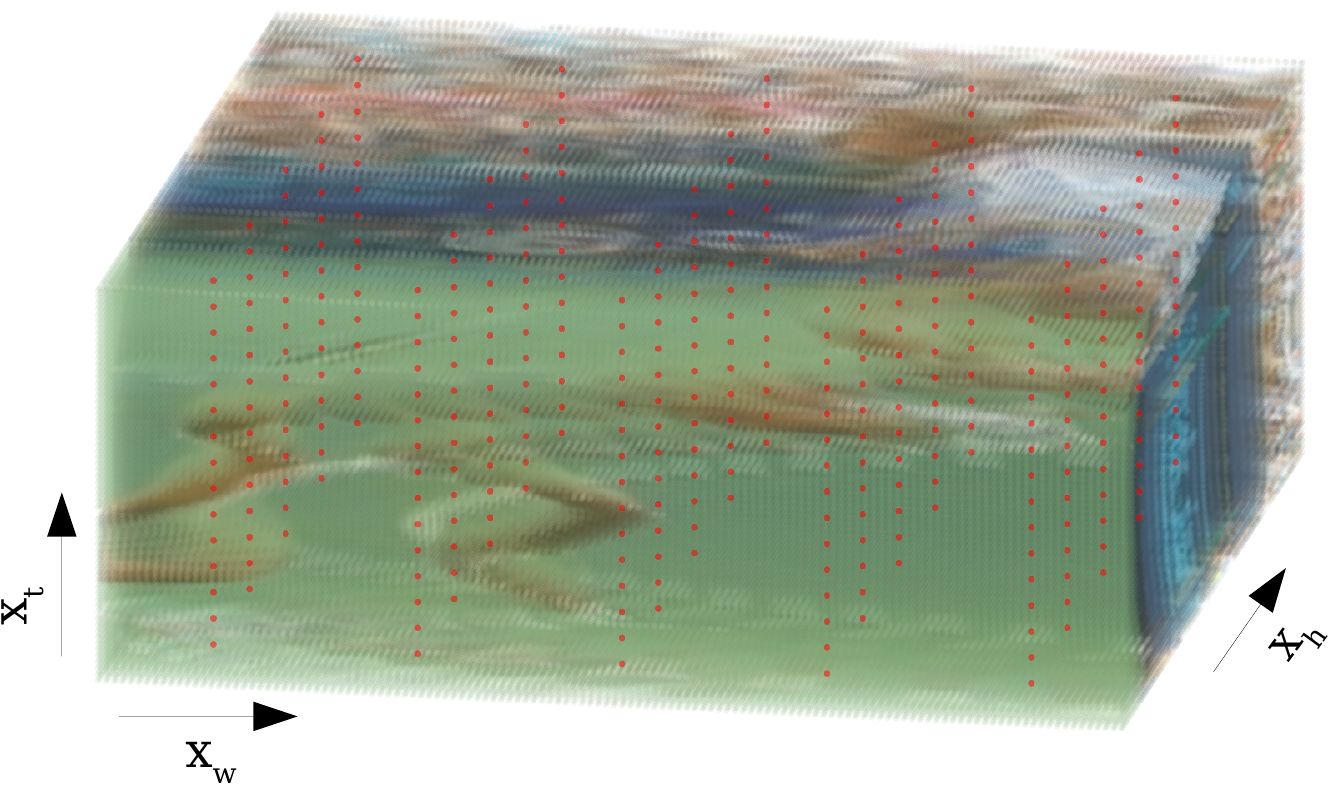};
\end{axis}

%

\end{tikzpicture}
\caption{Arrangement of data samples by stacking frames one after the other with regular grid of kernels as initialization drawn as red dots as their center positions.}
\label{fig:wo_mc_and_reg_kernel_grid}
\end{figure}

\begin{figure*}[tb]
\centering
\subfloat[Topdown View]{\begin{tikzpicture}

\begin{axis}[%
width=\scaleGMCplotW\textwidth,
height=\scaleGMCplotH\textwidth * 0.2632,
at={(0.0in,0.0in)},
scale only axis,
xmin=0,
xmax=1945,
y dir=reverse,
ymin=0,
ymax=512,
axis background/.style={fill=white},
axis line style={draw=none},
xticklabel=\empty,
yticklabel=\empty,
xtick=\empty,
ytick=\empty,
]
\addplot [forget plot] graphics 
[xmin=0,xmax=1945,ymin=0,ymax=512]{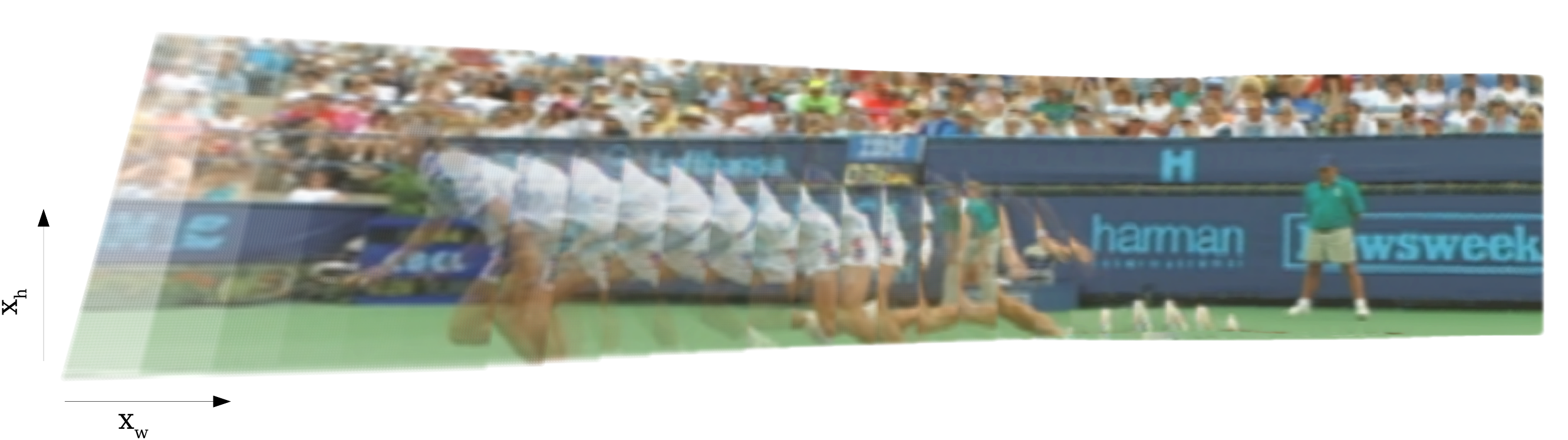};
\end{axis}

%

\end{tikzpicture}\label{subfig:topdown_stefan}}
\subfloat[Side View with illustrated initial center positions of kernels]{\begin{tikzpicture}

\begin{axis}[%
width=\scaleGMCplotW\textwidth,
height=\scaleGMCplotH\textwidth * 0.224557522,
at={(0.0in,0.0in)},
scale only axis,
xmin=0,
xmax=903,
y dir=reverse,
ymin=0,
ymax=294,
axis background/.style={fill=white},
axis line style={draw=none},
xticklabel=\empty,
yticklabel=\empty,
xtick=\empty,
ytick=\empty,
]
\addplot [forget plot] graphics 
[xmin=0,xmax=903,ymin=0,ymax=294]{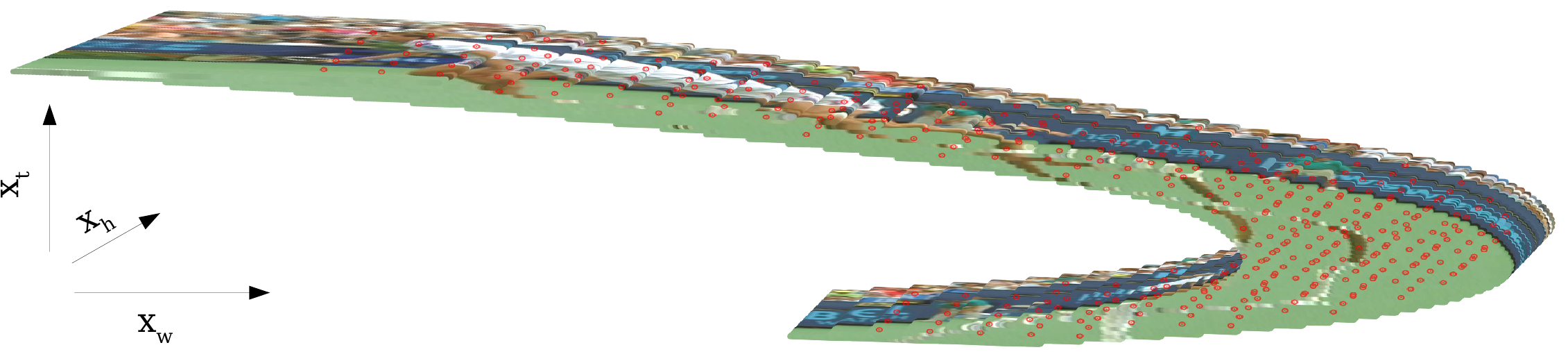};
\end{axis}

%

\end{tikzpicture} \label{subfig:side_stefan}}
\caption{Illustration of how the data samples are arranged after Global Motion Compensation}
\label{fig:GMC_on_data_samples}
\end{figure*}

\subsection{Parameter Optimization}
The model training follows \cite{tfsmoe} and \cite{dcc_smoe}. 
A multitask optimization technique is employed to train the parameters of the SMoE model.
As the SSIM metric is supposed to represent the human visual perception more precise than the Mean Squared Error (MSE) the main task is as follows:
\begin{align} \label{eq:loss_ssim}
    \mathcal{L}^\text{SSIM} := \left(  1 - \frac{1}{T}\sum_{t=1}^T \text{SSIM}(I_{t,\text{Target}}, I_{t,\text{Rec}}) \right)
\end{align}
where $I_{t,\text{Target}}$ and $I_{t,\text{Rec}}$ are the original and reconstructed images of frame $t$, respectively, and $T$ being the total number of frames. \\
In addition, a sparsity promoting regularization loss
\begin{align}
    \mathcal{L}^\text{S} := \lambda_\text{S} \cdot \sum\limits_{k=1}^K \pi_k
\end{align}
weighted by $\lambda_\text{S}$ is introduced to ensure that the underlying video is represented as sparse as possible while providing high reconstruction qualities. 
This $\ell_1$ regularization works by gradually decreasing the values of the mixing coefficients $\pi$ if the corresponding kernel has no contribution to the regression function, and thus, can be removed from the model.
The resulting loss function for minimization is composed as follows
%
\begin{align}
    \mathcal{L} := \mathcal{L}^\text{SSIM} + \mathcal{L}^\text{S} \text{.}
\end{align}
By following the negative gradient $-\nabla \mathcal{L}$ using Gradient Descent one can find a set of parameters $\mat{A}_k$, $\vec{m}_{0,k}$, $\vec{\mu}_k$ and $\pi_k$: \looseness=-1
\begin{align} \label{eq:argminL}
    \underset{\mat{A},\vec{m},\vec{\mu},\vec{\pi}}{\arg\min}\left\{ \mathcal{L} \right\} \text{.}
\end{align}
Note that the slopes in each channel $\mat{M}_k$ have been neglected in (\ref{eq:argminL}) and remain zero as they barely contribute to the regression quality as stated in \cite{dcc_smoe}.
%

%

\subsection{Parametric Motion Models}
Parametric Motion Models (PMM) are able to describe the displacement for each pixel at position $(x_w,x_h)^T$ to its new position $(x'_w, x'_h)^T$ within the subsequent frame by:
\begin{align}
w' \cdot
    \begin{pmatrix}
    x'_w \\
    x'_h \\
    1
    \end{pmatrix}
=
    \begin{pmatrix}  
        h_0 & h_1 & h_2 \\
        h_3 & h_4 & h_5 \\
        h_6 & h_7 & 1     
    \end{pmatrix}
\cdot
\begin{pmatrix}
    x_w \\ x_h \\ 1
\end{pmatrix} \label{eq:pmm}
\end{align}
including combinations of translation, zoom, rotation, shearing and perspective transformation \cite{tokPMM}.
Eq.\ (\ref{eq:pmm}) describes a homography with $p=8$ degrees of freedom and a scaling factor $w'=1/(h_6 \cdot x_w + h_7 \cdot x_h + 1)$.
The number of parameters of the homography can be reduced incrementally as follows:
\begin{itemize}
    \item $p=8$: perspective transformation as of Eq. \ref{eq:pmm}
    \item $p=6$: affine transformation with $h_6 = h_7 = 0$, all lines remain parallel
    \item $p=4$: similarity transformation with $h_3 = - h_1$ and $h_4 = h_0$, no shearing, no reflection, no scaling
    \item $p=2$: translation with $h_0 = h_4 = 1$ and $h_1 = h_3 = 0$, no rotation
\end{itemize}
%
The most accurate motion compensation can be achieved with $p=8$.
However, lower degrees of freedom may be desired in terms of rate-distortion performance for applications like video coding. The trade-off between the number of parameters $p$ and the associated gain in reconstruction quality is detailed in Sec.\ \ref{sec:experiments}. \looseness=-2 \\
%
The spatio-temporal arrangement of data samples within the continuous pixel domain of a video sequence without any motion compensation is depicted in Fig.\ \ref{fig:wo_mc_and_reg_kernel_grid}.
By having two spatial and one temporal dimensions each pixel location is composed by $\vec{x} = (x_w, x_h, x_t)^T$.
For illustration purpose we choose a crop of the well-known test video \emph{Stefan} seized by $128 \times 128$ and $64$ frames.
The camera is following the Tennis player resulting in non-linear global motions.
It can be seen that the frames are stacked by one after the other along the temporal dimension as in \cite{Videosmoe}.
Although the pixel domain is considered as continuous space the data samples are limited to discrete locations due to the temporal and spatial resolution of the source imagery.\\
For illustration purposes, the center positions of the kernels after initialization on a $5\times 5\times 15$ grid are depicted in Fig.\ \ref{fig:wo_mc_and_reg_kernel_grid}.\\
Fig.\ \ref{fig:GMC_on_data_samples} shows the positioning of the same data samples as in Fig.\ \ref{fig:wo_mc_and_reg_kernel_grid} after GMC of complexity $p=8$ has been applied.
Note that the GMC is performed on the pixel positions without the need to resample the image data.
Thus, the pixel amplitudes keep unmodified and no loss of information occurs.
The topdown view in Fig.\ \ref{subfig:topdown_stefan}
illustrates how the same content within the background is now spatially aligned along the temporal dimension.
We added transparency to each pixel to make the alignment visible.
Nevertheless, the motion of the foreground is easily recognizable. \\
The actual camera motion can be seen in Fig.\ \ref{subfig:side_stefan} as it is panning to right and then turns back to left.
Furthermore, Fig.\ \ref{subfig:side_stefan} depicts the distribution of the same kernels as of Fig.\ \ref{fig:wo_mc_and_reg_kernel_grid} after GMC has been applied.\\
By integrating GMC into SMoE framework the PMM parameter set $\mat{H} = [h_1, h_2, \ldots, h_8 ]$ can be added to the optimization problem and Eq.\ \ref{eq:argminL} needs to be rewritten as:
\begin{align} \label{eq:argminL_w_PMM}
    \underset{\mat{H},\mat{A},\vec{m},\vec{\mu},\vec{\pi}}{\arg\min}\left\{ \mathcal{L} \right\} \text{.}
\end{align}
This renders the approach more robust against noisy or sub-optimal initial estimates of the PMM by fine-tuning them directly towards better reconstruction qualities. 

\begin{figure*}[tb]
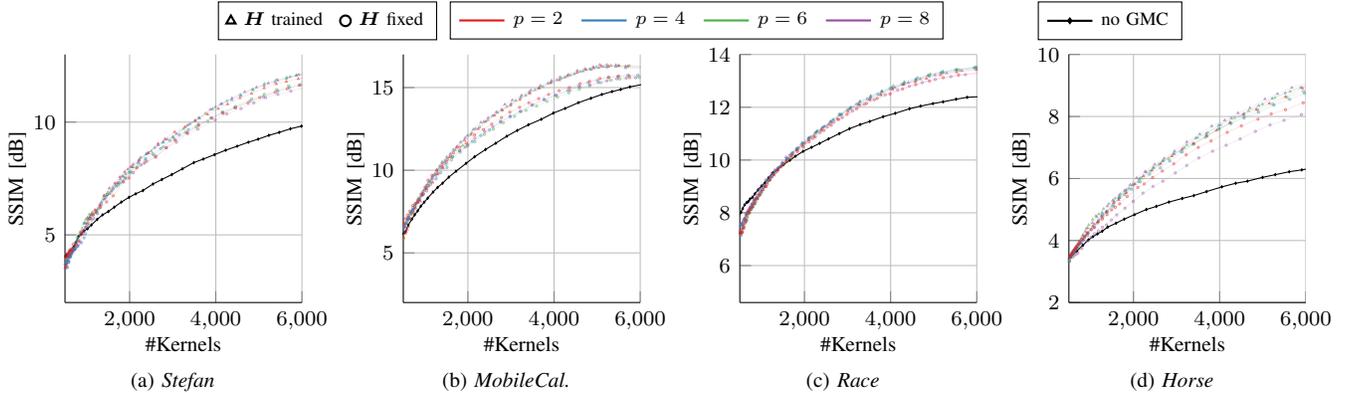

\centering
\hspace{0.8cm}\subfloat{
\definecolor{mycolor1}{rgb}{0.894117647,0.101960784,0.109803922}
\definecolor{mycolor2}{rgb}{0.215686275,0.494117647,0.721568627}
\definecolor{mycolor3}{rgb}{0.301960784,0.68627451,0.290196078}
\definecolor{mycolor4}{rgb}{0.596078431,0.305882353,0.639215686}

\definecolor{schwarz}{rgb}{0,0,0}%

\def\shiftright{.06in}

\begin{tikzpicture}

\begin{axis}[%
width=0\textwidth,
height=0\textwidth,
at={(0.25in,\tikzlegensPosY)},
scale only axis,
xmin=0.0,
xmax=0.1,
xlabel style={font=\color{white!15!black}},
xlabel={bitrate [bpp]},
ymin=0,
ymax=1,
ylabel style={font=\color{white!15!black}},
ylabel={PSNR [dB]},
minor tick num = 1,
legend columns=-1,
ylabel near ticks,
axis x line*=bottom,
axis y line*=left,
legend style={legend cell align=left, align=left, /tikz/every even column/.append style={column sep=0.16cm},font=\fontsize{7}{\baselineskip}\selectfont, column sep=1pt}
]

\addplot [color=schwarz, mark=diamond, thick, mark options={solid, schwarz,scale=\scaleMarks*2}]
  table[row sep=crcr]{%
0.15875244140625	28.1299064315678\\
};
\addlegendentry{no GMC}

\end{axis}

\begin{axis}[%
width=0\textwidth,
height=0\textwidth,
at={(-1in,\tikzlegensPosY)},
scale only axis,
xmin=0.0,
xmax=0.1,
xlabel style={font=\color{white!15!black}},
xlabel={bitrate [bpp]},
ymin=0,
ymax=1,
ylabel style={font=\color{white!15!black}},
ylabel={PSNR [dB]},
minor tick num = 1,
legend columns=-1,
ylabel near ticks,
axis x line*=bottom,
axis y line*=left,
legend style={legend cell align=left, align=left, /tikz/every even column/.append style={column sep=0.16cm},font=\fontsize{7}{\baselineskip}\selectfont, column sep=1pt}
]


\addplot [color=mycolor1, mark=none, thick]
  table[row sep=crcr]{%
0.15875244140625	28.1299064315678\\
};
\addlegendentry{$p=2$}

\addplot [color=mycolor2, mark=none, thick]
  table[row sep=crcr]{%
0.15875244140625	28.1299064315678\\
};
\addlegendentry{$p=4$}

\addplot [color=mycolor3, mark=none, thick]
  table[row sep=crcr]{%
0.15875244140625	28.1299064315678\\
};
\addlegendentry{$p=6$}

\addplot [color=mycolor4, mark=none, thick]
  table[row sep=crcr]{%
0.15875244140625	28.1299064315678\\
};
\addlegendentry{$p=8$}

\end{axis}

\begin{axis}[%
width=0\textwidth,
height=0\textwidth,
at={(-3.65in,\tikzlegensPosY)},
scale only axis,
xmin=0.0,
xmax=0.1,
xlabel style={font=\color{white!15!black}},
xlabel={bitrate [bpp]},
ymin=0,
ymax=1,
ylabel style={font=\color{white!15!black}},
ylabel={PSNR [dB]},
minor tick num = 1,
legend columns=-1,
ylabel near ticks,
axis x line*=bottom,
axis y line*=left,
legend style={legend cell align=left, align=left, /tikz/every even column/.append style={column sep=0.16cm},font=\fontsize{7}{\baselineskip}\selectfont, column sep=1pt}
]
\addplot [color=schwarz, only marks, thick, mark=triangle]
  table[row sep=crcr]{%
0.15875244140625	28.1299064315678\\
};
\addlegendentry{$\mat{H}$ trained}

\addplot [color=schwarz, only marks, thick, mark=o,solid]
  table[row sep=crcr]{%
0.15875244140625	28.1299064315678\\
};
\addlegendentry{$\mat{H}$ fixed}
\end{axis}

\end{tikzpicture}} \\
\vspace{-0.37cm}
\renewcommand{\thesubfigure}{a}\subfloat[\emph{Stefan}]{\input{fig/stefan} \label{subfig:stefan_results}}
\hspace{-0.1cm}%
\renewcommand{\thesubfigure}{b}\subfloat[\emph{MobileCal.}]{\input{fig/mobile}} 
\renewcommand{\thesubfigure}{c}\subfloat[\emph{Race}]{\input{fig/race}}\hspace{-0.1cm}%
\renewcommand{\thesubfigure}{d}\subfloat[\emph{Horse}]{\input{fig/horse}}
\caption{Comparison of the reconstruction quality of obtained SMoE models learned using different initialization setups and GMC complexity $p$ in terms of SSIM and number of used kernels. Note that the SSIM metric is converted to dB to make differences better visible ($\text{SSIM}_\text{dB} = - 10 \cdot log_{10}(1-d)$ with $d$ being the SSIM value; best viewed in color and zoomed-in)}
\label{fig:results}
\end{figure*}

\section{Experiments}
\label{sec:experiments}
In this section we evaluate the performance of the proposed method for modelling sparse representations of video data and analyze the impact of the application of GMC.
The implementation is based on the framework of \cite{tfsmoe} and follows its Adam based optimization scheme.\\
For the application to color video we have to choose deviating learning rates of $1$, $10^{-4}$ for $\mat{A}$, $\vec{\pi}$, respectively, and $10^{-3}$ for $\vec{\mu}$, $\vec{m}, \mat{H}$ as their corresponding gradients are of different magnitudes.\\
The modeling process of the underlying video data contains four steps:
First, Global Motion Estimation is done according to the chosen complexity $p$ to initialize the respective parameters in $\mat{H}$.
Then, the model is initialized and pre-trained without regularization for sparsification. 
Following, the model is sparsified to the desired degree by introducing $\lambda_S > 0$.
Finally, a fine-tuning step follows to maximize the reconstruction quality while no regularization is applied. 
Note that kernels will be removed from the model during all training steps if their corresponding mixing coefficients reach $\pi_k \leq 0$.

\textbf{Global Motion Estimation} is performed to initialize parameters of the motion model of the proposed method.
The necessary point correspondences between subsequent frames are obtained by means of optical flow \cite{raft_optflow}.
We choose to sample these motion vectors on a regular grid to capture the global motion of the entire frame instead of feature based sampling strategies that tend to focus on foreground or texture-rich objects. 
Finally, the parameters of the motion model are computed using a RANSAC-based \cite{ransac} approach and by minimizing the re-projection error of the inliers.

\textbf{Initialization and Pre-Training} Initialization is done before GMC is applied.
The center positions $\vec{\mu}$ of the kernels are distributed on a regular grid by $k_w \times k_h \times k_t$ specifying the amount of kernels per respective dimension (see Fig.\ \ref{fig:wo_mc_and_reg_kernel_grid}). 
The bandwidths $\mat{A}$ are initialized such that the distance between the centers of two neighboring kernels equals two standard deviations $2\sigma$ regarding the corresponding dimension. 
The mixing coefficients are all set to $\pi_k = 1$. 
Then, the GMC is applied to the pixel locations as well as to the center positions (see Fig. \ref{subfig:side_stefan}). Note that the displacement of kernels by GMC is only done in the initialization step to find their start positions. 
During training merely the pixel locations are affected by the GMC.\\
Subsequently, the offsets $\vec{m}_{0,k}$ are initialized by averaging each of the respective channels (Y,U,V) of the training data where the gating function of the corresponding kernel has maximum influence. 
Finally, the pre-training is employed for 10k iterations with no sparsification applied ($\lambda_\text{S} = 0$).

\textbf{Regularization} for sparsification is performed after pre-training.
According to \cite{tfsmoe} best results are achieved by slowly introducing the regularization term and exponentially increasing the coefficient $\lambda_\text{S}$. 
We choose a schedule as follows: $\lambda_\text{S} = s^2/(k_w \cdot k_h \cdot k_t)$ with $s$ evenly distributed in $[1, 150]$ over 50 steps.
To consider different sensitivity levels depending on the initialization of kernels the normalization with the fixed number of initial kernels $k_w \cdot k_h \cdot k_t$ is incorporated.
In each step the model is trained for 1k iterations.\\
\textbf{Fine-Tuning}
is done to quickly recover from the previous trade-off between the SSIM $\mathcal{L}^\text{SSIM}$ and the sparsification loss $\mathcal{L}^\text{S}$ towards better reconstruction quality. 
This step is employed for 200 training iterations with $\lambda_\text{S} = 0$.

\subsection{Results}
We evaluated the proposed method on various small test sequences of size $128   \times  128$ and $64$ frames due to high complexity reasons. 
Therefore, crops from the well-known sequences \emph{Mobile\&Calendar}, \emph{RaceHorse}, \emph{Race}, and \emph{Stefan} have been investigated. 
As the luminance channel is the most important regarding the human visual system compared to the chrominance channels the SSIM function in Eq.\ \ref{eq:loss_ssim} is evaluated for each channel independently and summed with weights $\text{Y:U:V} \leftrightarrow 6\text{:}1\text{:}1$ to give the luminance channel the necessary relevance within the optimization process. 
The same weights are used to determine the SSIM value for validation.\\
We choose an initial kernel grid of $32 \! \! \times \! \! 32 \! \! \times \! \! 8$ to have the same kernel density along the spatial dimensions as in \cite{tfsmoe}. 
As we expect more redundancies between adjacent frames we set $k_t = 8$.
In addition to that, we tested each setup with GMC of complexity $p = \{2,4,6,8\}$ and without any motion compensation for comparison. 
The training was performed with and without tuning the GMC parameters $\mat{H}$ after initial estimation to investigate if the joint optimization of the SMoE Model and GMC parameters offers additional gains towards better reconstruction qualities.

\begin{table}[t]
\centering
\scriptsize
\caption{Quality examples in terms of PSNR in dB and SSIM for $K \approx 5000$ and various PMM complexity $p$ per model.}
\label{tab:results}
\setlength{\tabcolsep}{1.5pt}
\begin{tabular}{l|c>{\centering}p{9mm}|p{4mm}cccc}
\hline
\hline
\multicolumn{8}{c}{}\\[-6pt]
Sequence & \multicolumn{2}{c|}{no GMC} && \multicolumn{2}{c}{$\mat{H}$ fixed} & \multicolumn{2}{c}{$\mat{H}$ trained} \\[0pt]
                 & PSNR [dB]              & SSIM                    & $p$ & PSNR [dB] & SSIM      & PSNR [dB] & SSIM\\[1.5pt]
 Stefan          &\multirow{1}{*}{$28.28$}&\multirow{1}{*}{$0.8810$}& $2$ & $30.60$   &  $0.9222$ & $31.07$   &  $0.9294$ \\
                 &                        &                         & $4$ & $30.56$   &  $0.9228$ & $31.21$   &  $0.9312$ \\
                 &                        &                         & $6$ & $30.77$   &  $0.9248$ & $31.10$   &  $0.9286$ \\
                 &                        &                         & $8$ & $30.34$   & $0.9178$  & $31.25$   &  $\boldsymbol{0.9319}$ \\[1.5pt]
MobileCal.       &\multirow{1}{*}{$31.65$}&\multirow{1}{*}{$0.9652$}& $2$ & $32.54$   & $0.9721$  & $33.12$   & $0.9759$ \\
                 &                        &                         & $4$ & $32.49$   & $0.9714$  & $33.12$   & $0.9760$ \\
                 &                        &                         & $6$ & $32.33$   & $0.9704$  & $33.22$   & $0.9762$ \\
                 &                        &                         & $8$ & $32.35$   & $0.9706$  & $33.31$   & $\boldsymbol{0.9770}$ \\[1.5pt]
Race             &\multirow{1}{*}{$33.77$}&\multirow{1}{*}{$0.9289$}& $2$ & $34.00$   & $0.9498$  & $34.22$   & $0.9524$  \\
                 &                        &                         & $4$ & $34.26$   & $0.9523$  & $34.23$   & $0.9522$  \\
                 &                        &                         & $6$ & $34.19$   & $0.9520$  & $34.34$   & $\boldsymbol{0.9528}$  \\
                 &                        &                         & $8$ & $34.29$   & $0.9505$  & $34.37$   & $0.9523$ \\[1.5pt]
RaceHorse        &\multirow{1}{*}{$23.00$}&\multirow{1}{*}{$0.7509$}& $2$ & $25.57$   & $0.8384$  & $26.14$   & $0.8498$  \\
                 &                        &                         & $4$ & $25.92$   & $0.8464$  & $26.28$   & $0.8538$  \\
                 &                        &                         & $6$ & $26.01$   & $0.8488$  & $26.32$   & $0.8554$  \\
                 &                        &                         & $8$ & $25.03$   & $0.8230$  & $26.47$   & $\boldsymbol{0.8608}$  \\
\multicolumn{8}{c}{}\\[-6pt]
\hline\hline
\end{tabular}
\end{table}

%
%
%
Fig.\ \ref{fig:results} and Tab.\ \ref{tab:results} show the achieved reconstruction results for the four aforementioned test sequences. The vanilla approach in which subsequent frames are merely stacked one after the other is declared as "no GMC". 
The results of our proposed method are depicted in different colors depending of the complexity $p$. Furthermore, we separated the curves by either a circle or a triangle stating $\mat{H}$ has been fixed or further optimized, respectively.
Although the aforementioned sparsification procedure provides the best results in general, unfortunately, this is not the case for \emph{MobileCal} for $K\!>\!5600$.
As the reconstruction error after pretraining is very small, the sparsification introduces a high relative drop in quality. This may be compensated by additional training iterations.
However, the quality for numbers of kernels in the interval $[ 2000, 5000 ]$ are of most interest, which are not affected by this behavior.
It can be seen that our method outperforms the vanilla approach for any complexity $p$ and number of kernels $K\! >\! 2000$.
Besides, by training $\mat{H}$ we can reach additional significant gain in terms of SSIM as shown in Tab.\ \ref{tab:results} by up to $0.0378$.
It turns out that the overall best results are achieved by using GMC of complexity $p=8$ with trained $\mat{H}$ except for \emph{Race}. 
The global motion in this scene consists only of panning of the camera, i.e. translatory motion that can be described with $p=2$ sufficiently. Therefore, the results for any $p\!>\!0$ are very close with the highest SSIM for $p\!=\!6$ at a low margin.
Due to highly nonlinear camera motions in \emph{Stefan} and \emph{Horse} very high gains can be achieved.
This is also illustrated in Fig.\ \ref{fig:visual_comparison}.
With no GMC textures are blurred as more kernels are required along the temporal domain whereas with our proposed approach the temporal complexity is greatly reduced. 
This allows to reallocate kernels spatially leading to the representation of finer details.\looseness=-1

\section{Summary and Conclusion}
A novel optimization method for video representation using SMoE was presented in this paper. By applying GMC on the pixel locations on each frame significant gains in sparsity can be achieved by reducing the amount of kernels up to $65.86\%$ compared to models with no GMC at same regression quality. We have shown that spatial alignment along the time dimension enables to exploit more redundancies within video data resulting in more detailed and global motion consistent reconstructions.
Future work will examine how this promising performance gain can be used to encode video data. In addition, the expansion of the  GMC towards light field videos will be investigated.




\bibliographystyle{unsrt}
\bibliographystyle{IEEEbib}
\bibliography{biblio}

\end{document}